\documentclass[11pt]{article}
\usepackage{times}
\usepackage{geometry}
\usepackage[utf8]{inputenc}
\usepackage{enumitem,amssymb}
\usepackage{ragged2e}
\newlist{thematic}{itemize}{8}
\setlist[thematic]{label=$\square$}
\usepackage{pifont}

\newcommand{\myvspace}{\vspace{0.1cm}}
\usepackage{enumitem}
\setlist[itemize]{leftmargin=*}
\usepackage{fancyhdr}
\usepackage[USenglish]{babel}
\usepackage[utf8]{inputenc}
\usepackage[T1]{fontenc}
\usepackage{epsfig}
\usepackage{wrapfig}
\usepackage{color}
\usepackage[vflt]{floatflt}
\usepackage{longtable}
\usepackage{caption}
\usepackage{jheppub}   
\usepackage[dvipsnames,svgnames,x11names]{xcolor}
\usepackage[all]{hypcap} 
\usepackage{amssymb,amsmath}
\usepackage{longtable}
\usepackage{amssymb}
\usepackage{amsmath, xspace}
\usepackage{graphics,graphicx} 
\usepackage{rotating}
\usepackage{color}
\usepackage{xcolor}
\usepackage{multirow}
\usepackage{subfigure}
\usepackage{sectsty}
\usepackage[outercaption]{sidecap}
\sidecaptionvpos{figure}{c}

\usepackage{eso-pic,graphicx}
\usepackage{tikz}

\usepackage{xcolor}
\definecolor{cobalt}{rgb}{0., 0.35, 0.56}
\usepackage{hyperref}

\AddToShipoutPictureBG{%
  \ifnum\value{page}=1
    \AtPageLowerLeft{%
      \includegraphics[width=\paperwidth]{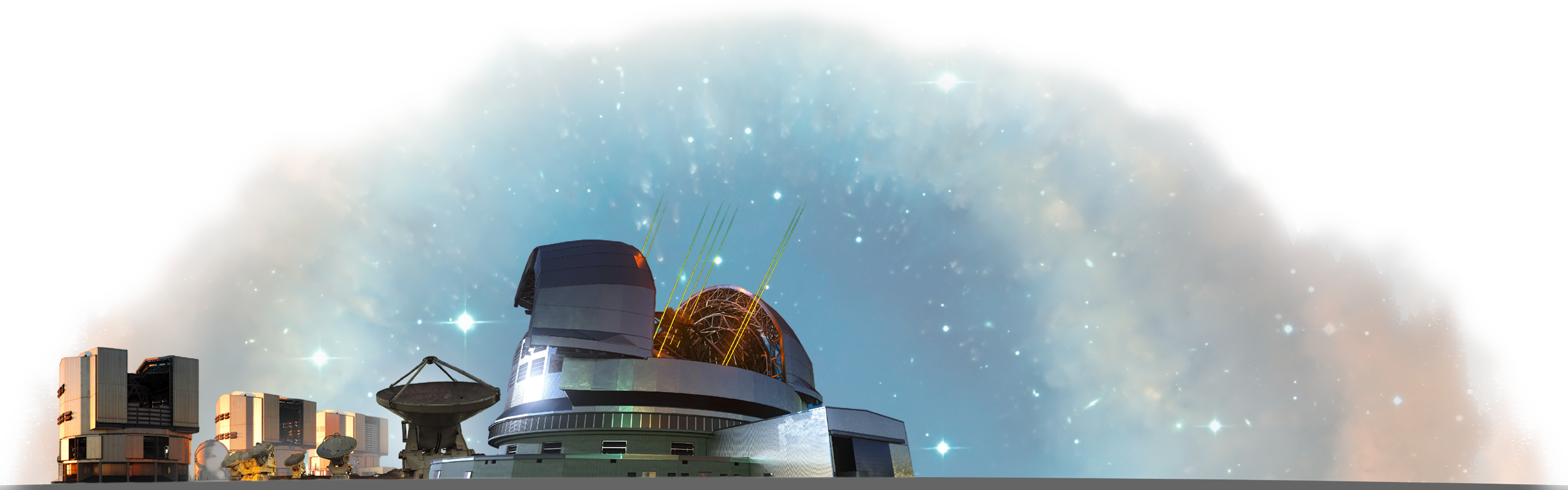}%
    }%
  \fi
}

\usepackage{enumitem}
\setlist[enumerate]{itemsep=0pt, parsep=0pt}

\usepackage{natbib}

\definecolor{DarkGreen}{rgb}{0.0, 0.3, 0.0}
\definecolor{purple}{rgb}{0.5, 0.0, 0.5}
\definecolor{red}{rgb}{1, 0.0, 0.0}
\definecolor{green}{rgb}{0, 1.0, 0.0}

\def\3he{$^3{\rm He}$}

\def\lsim{\mathrel{\lower2.5pt\vbox{\lineskip=0pt\baselineskip=0pt
           \hbox{$<$}\hbox{$\sim$}}}}

\def\gsim{\mathrel{\lower2.5pt\vbox{\lineskip=0pt\baselineskip=0pt
           \hbox{$>$}\hbox{$\sim$}}}}

\begin{document}

\raggedright
\Large
ESO Expanding Horizons initiative 2025 \linebreak
Call for White Papers

\vspace{1.cm}

\raggedright
\huge

Toward the time-domain spectroscopic study of the dynamic life of stars: from accretion to magnetic activity


\bigskip
\normalsize

\raggedright
\Large
\textbf{Scientific Question:} How can the next big ESO telescope revolutionize our understanding of stellar dynamical processes in young stars
by monitoring them
at baselines from tens of minutes to tens of years? 
\linebreak
\normalsize

\myvspace

\textbf{Authors:} 
Fatemeh Zahra Majidi (fatemeh.majidi@inaf.it, INAF OACN, Italy); Amelia Bayo (ESO-Garching, Germany); Marc Audard (University of Geneva, Switzerland); Francisco Jos\'e Galindo-Guil (CEFCA, Spain); Rosaria Bonito (INAF - Osservatorio Astronomico di Palermo, Italy);

\textbf{Endorsers:} Katia Biazzo (INAF - Osservatorio Astronomico di Roma, Italy); Loredana Prisinzano (INAF - Osservatorio Astronomico di Palermo, Italy); Mario Giuseppe Guarcello (INAF - Osservatorio Astronomico di Palermo, Italy); Eleonora Fiorellino (University of Bologna, Italy); Innocenza Busa (INAF OACT, Italy); Germano Sacco (INAF-Osservatorio Astrofisico di Arcetri); Richard I. Anderson (EPFL, Switzerland); Avraham Binnenfeld (EPFL, Switzerland); David Montes 
(Universidad Complutense de Madrid and IPARCOS-UCM, Spain)

\myvspace

\textbf{Science Keywords:} 
young stellar objects, star-forming regions, stellar magnetic activity

\begin{figure}[!ht]
    \centering
    \includegraphics[width=0.5\linewidth]{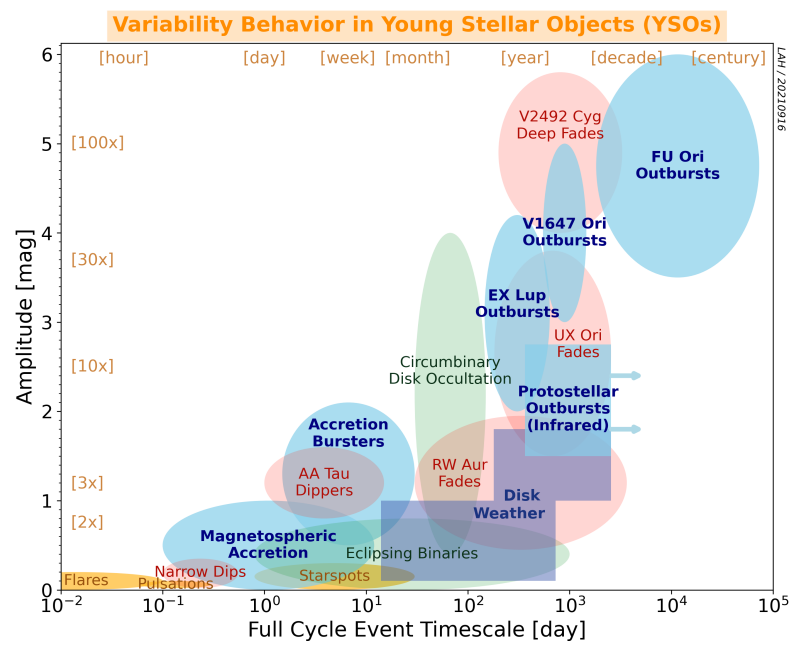}
    \caption*{\small Amplitude versus timescale for different types of YSO variability -- adapted from Fischer et al. 2023 \citep{fischer2023}.}
    \label{fig:placeholder}
\end{figure}

\newpage

\textbf{Abstract:} Stars and planets can be seen as the second fundamental building blocks of baryons in the universe (only second to the dust and gas in molecular clouds). Their formation involves dust grain growth of many orders of magnitude and a myriad of processes operating at time scales from a few tens to millions of years. Thus, investigating the formation and evolution of young stellar objects (YSOs) is of great importance in modern astronomy. Addressing this goal requires overcoming long-standing challenges in characterizing multifaceted phenomena that span a broad range of astrophysical processes—from protoplanetary disk evolution and planet formation to accretion dynamics and transient stellar events. Also, YSOs are complex systems that consist of several components: a central forming object, surrounded by a medium or disk from which the accretion process is at work, supersonic ejection of plasma in the form of collimated bipolar jets—which interact with the ambient medium through which they propagate—and all these components emit in a wide range of wavelengths. A facility capable of simultaneously tackling these diverse questions must deliver long-term, high-cadence spectroscopic monitoring of YSOs over time spans of at least a decade; especially because accretion/ejection processes in YSOs are characterized by a wide range of temporal variability: from short-term (hours-days) to long-term (months-years) variability due to rotation, accretion, magnetic activity, etc. Such a mission demands a spectroscopic platform considering a solid time-domain astronomy framework, providing repeated observations over wide fields and supporting multiple cadence strategies tailored to distinct scientific objectives. 

\begin{itemize}
 
\item \textbf{Science Drivers}

\end{itemize}

Understanding the formation and early evolution of stars requires capturing a continuum of physical processes that unfold across spatial scales spanning more than six orders of magnitude and over timescales from minutes to decades. From the late-stage inflow of molecular clouds into filaments and clumps \citep{a_hacar2017b}, through the collapse of dense cores \citep{p_andre2014, m_tafalla2015b}, and the formation of protostellar disks, to the energetic events—accretion bursts, outflows, coronal mass ejections (CMEs), and magnetic flares—that dominate the early stellar environment, nearly every step in this chain is inherently time-variable. Yet despite decades of progress, our understanding of stellar formation and evolution very much relies on snapshot observations, which may be infrequently repeated, thus providing us access to specific conditions prevailing during the observations that may not represent the underlying intrinsic variability of processes occurring at different timescales. Despite the long timescale of star formation and evolution (on a human lifetime), these studies provide an incomplete and biased view of the physical processes by extrapolating mean properties (star formation rates, mass accretion rates, flare frequencies) from these rare observations.  \myvspace

A transformative advance requires frequent time-resolved, spectrally resolved access to the kinematics, radiative processes, and multi-component line tracers that encode the physics of the star-formation cycle. Wide-field photometric time-domain surveys \citep{a_siciliaaguilar2024} (such as Vera Rubin/LSST and Gaia) will have delivered exquisite temporal sampling and variability diagnostics but cannot provide the physical interpretation of those signals without corresponding spectroscopic information. To unlock this potential, a large ESO spectroscopic facility equipped with both integral field spectroscopy (IFS) to probe the complex surrounding environment of YSOs and multi-object spectroscopy (MOS) operating at high resolution (R$\sim$40,000) and low resolution (R$\sim$4,000) over a wide field of view (to scan crowded star forming regions homogeneously) enables a comprehensive mapping of the relevant dynamic processes that shape YSOs.

\begin{itemize}
 
\item \textbf{Tracing the Full Hierarchy of Star Formation: From Cloud Inflows to Disk Microphysics}
\end{itemize}

At the largest scales ($\sim$1–10 pc), low-resolution IFS capability would map the velocity structure and physical conditions of inflowing material along the filaments feeding protostellar clusters. Spectrally resolved line ratios of H I, H$_2$, and key forbidden tracers would probe turbulence dissipation, shock heating, and the impact of stellar feedback on cloud assembly. While structural evolution on parsec scales occurs over $10^{5}$–$10^{6}$ yr, long-term monitoring of the embedded sources within these filaments, using diagnostics such as CO bandheads and near-infrared permitted lines, would capture variability in their immediate ($\lesssim$1 au) environments, linking accretion changes to local heating and feedback processes. This capability provides a critical multi-scale context for understanding how large-scale flows regulate the earliest phases of star and cluster formation.\myvspace

On intermediate scales ($\sim$10$^{2}$–10$^{3}$ au), MOS observations with moderate-to-high spectral resolution would trace protostellar envelopes, wide-angle winds, and collimated jets, enabling time-resolved measurements of mass-loss rates, jet velocities, and variability in forbidden-line emission. Monitoring these outflows on monthly-to-yearly timescales is essential for identifying episodic ejection events and linking them to accretion variability in the inner disk. In particular, the formation and propagation of shock-excited knots provide direct signatures of discrete accretion bursts and offer crucial constraints on the jet-disk coupling mechanism. This intermediate regime thus bridges cloud dynamics with disk physics and connects environmental conditions with the early evolution of angular momentum and magnetic fields\citep{a_siciliaaguilar2020b}.\myvspace

At the $\sim$100 au scale, corresponding to the outer regions of protoplanetary disks, moderate-to-high resolution spectroscopy can follow changes in extended disk winds and scattered-line emission, but the physical processes governing accretion originate much closer in. High/Low-resolution (R$\sim40{,}000 \hspace{0.2cm} \& \sim4{,}000$) spectroscopy of diagnostic lines (for monitoring accretion funnels, jets, disk winds, and flares to resolve multi-component line profiles in H I, He I, Ca II, [O I], [S II], CO rovibrational bands) probes sub-au to few-au regions, allowing decomposition of the emission and absorption components arising from accretion shocks, magnetospheric funnel flows, disk winds, and turbulence in the hot inner disk. Nightly-to-monthly cadence spectroscopy captures classical low-level accretion flickering, structural evolution of the inner disk \citep{duy_cuong_nguyen2009a}, transitions between accretion states \citep{bayo2012, stauffer2014}, and magnetospheric reconnection events \citep{barrado2002a}. Such time-series observations are indispensable for characterizing the interplay between disk evolution, magnetic topology, and variable mass transport through the star–disk interface.\myvspace 

At the smallest spatial scales (few au down to stellar radii), high-cadence MOS and IFS spectroscopy would capture the most rapid and energetic processes: accretion bursts \citep{a_kospal2011b, m_audard2014, a_siciliaaguilar2012}, magnetic flares, CMEs, and stellar wind variability \citep{suzan_edwards2003, christopher_m_johnskrull_2007a, justyn_campbellwhite2023b}. These phenomena occur on timescales of minutes to hours, driving instantaneous changes in line profiles that encode the geometry and energetics of magnetospheric flows. High-cadence spectroscopy is uniquely capable of resolving the velocity structure of flare-driven winds, quantifying CME masses, and constraining how rapid accretion fluctuations propagate into ejection events — not feasible through photometry. Capturing this short-timescale behavior is essential for understanding the physics of the accretion–ejection connection, the launching of jets and outflows, and the resulting impacts on disk ionization, chemistry, the early conditions for planet formation, and ultimately the habitability of emerging planetary systems. \myvspace

\begin{itemize}
    
\item \textbf{Motivation: A Spectrally Time-Resolved Map of Physical Processes}

\end{itemize}

The major motivation behind this paper is addressing the science drivers above, generating a comprehensive mapping of physical processes, built not from static, population-averaged indicators, but from time-resolved, spectrally resolved observations across thousands of YSOs in a homogeneous fashion. With decade-long monitoring and multi-resolution spectroscopy, it will be possible to systematically connect observables(variability amplitudes, velocity components, excitation conditions) to concrete physical mechanisms such as angular-momentum transport, mass accretion rate, magnetic reconnection, shock dissipation, or infall–outflow coupling \citep{a_siciliaaguilar2020b}.\myvspace

In effect, the facility would build the first dynamical, multi-scale, empirical atlas of star formation: linking cloud-scale inflow variability, to envelope and jet evolution, to disk accretion physics, to magnetic stellar activity. This dataset would redefine our understanding of early stellar evolution and the physical boundary conditions of planet formation.

\begin{itemize}
 
\item \textbf{Key Questions}

\end{itemize}

Here are the major open questions that the next big telescope developed by ESO will address through conducting a decadal spectroscopic survey of YSOs:

\begin{itemize}[itemsep=0pt, parsep=0pt]

\item How does the inner disk distribute and modulate angular momentum on timescales from hours to years?

\item Are variability patterns on au scales causally linked to fluctuations in large-scale inflow?

\item How do inflow fluctuations correlate with episodic accretion events onto individual protostars?

\item How do jet velocities and mass-loss rates vary on short (daily) and long (annual–decadal) timescales?

\item What are the plasma temperatures, excitation properties, and velocity-resolved kinematic signatures of flares, CMEs, and magnetic reconnection events in pre-main-sequence stars?

\item How do these events influence disk chemistry, ionization structure, and the early atmospheric evolution of close-in forming planets? 

\item How to unravel the temporal dependence of the accretion rate on stellar age, and distinguish it from the effects of stellar mass and observational biases.

\end{itemize}


\bibliographystyle{plain} 
\bibliography{references}

@ARTICLE{fischer2023,
       author = {{Fischer}, W.~J. and {Hillenbrand}, L.~A. and {Herczeg}, G.~J. and {Johnstone}, D. and {Kospal}, A. and {Dunham}, M.~M.},
        title = "{Accretion Variability as a Guide to Stellar Mass Assembly}",
    booktitle = {Protostars and Planets VII},
         year = 2023,
       editor = {{Inutsuka}, S. and {Aikawa}, Y. and {Muto}, T. and {Tomida}, K. and {Tamura}, M.},
       series = {Astronomical Society of the Pacific Conference Series},
       volume = {534},
        month = jul,
        pages = {355},
          doi = {10.48550/arXiv.2203.11257},
archivePrefix = {arXiv},
       eprint = {2203.11257},
 primaryClass = {astro-ph.SR},
       adsurl = {https://ui.adsabs.harvard.edu/abs/2023ASPC..534..355F}
}

@ARTICLE{a_siciliaaguilar2024,
       author = {{Sicilia-Aguilar}, A. and {Kahar}, R.~S. and {Pelayo-Bald{\'a}rrago}, M.~E. and {et al.}},
        title = "{North-PHASE: studying periodicity, hot spots, accretion stability, and early evolution in young stars in the Northern hemisphere}",
      journal = {\mnras},
         year = 2024,
        month = aug,
       volume = {532},
       number = {2},
        pages = {2108-2132},
          doi = {10.1093/mnras/stae1588},
archivePrefix = {arXiv},
       eprint = {2406.16702},
 primaryClass = {astro-ph.SR},
       adsurl = {https://ui.adsabs.harvard.edu/abs/2024MNRAS.532.2108S}
}

@ARTICLE{duy_cuong_nguyen2009a,
       author = {{Nguyen}, D.C. and {Jayawardhana}, R. and {van Kerkwijk}, M. H. and {et al.}},
        title = "{Disk Braking in young Stars: Probing Rotation in Chamaeleon i and Taurus-Auriga}",
      journal = {\apj},
         year = 2009,
        month = apr,
       volume = {695},
       number = {2},
        pages = {1648-1656},
          doi = {10.1088/0004-637X/695/2/1648},
archivePrefix = {arXiv},
       eprint = {0902.0001},
 primaryClass = {astro-ph.SR},
       adsurl = {https://ui.adsabs.harvard.edu/abs/2009ApJ...695.1648N}
}

@ARTICLE{a_kospal2011b,
       author = {{K{\'o}sp{\'a}l}, {\'A}. and {{\'A}brah{\'a}m}, P. and {Goto}, M. and {et al.}},
      journal = {\apj},
         year = 2011,
       volume = {736},
       number = {1},
        pages = {72},
          url = {https://iopscience.iop.org/article/10.1088/0004-637X/736/1/72}, 
}

@INPROCEEDINGS{m_audard2014,
       author = {{Audard}, M. and {{\'A}brah{\'a}m}, P. and {Dunham}, M.~M. and {et al.}},
        title = "{Episodic Accretion in Young Stars}",
    booktitle = {Protostars and Planets VI},
         year = 2014,
       editor = {{Beuther}, H. and {Klessen}, R.S. and {Dullemond}, C.P. and {Henning}, T.},
        month = jan,
        pages = {387-410},
          doi = {10.2458/azu_uapress_9780816531240-ch017},
archivePrefix = {arXiv},
       eprint = {1401.3368},
 primaryClass = {astro-ph.SR},
       adsurl = {https://ui.adsabs.harvard.edu/abs/2014prpl.conf..387A}
}

@ARTICLE{christopher_m_johnskrull_2007a,
       author = {{Johns-Krull}, Christopher M. and {Herczeg}, Gregory J.},
        title = "{How Hot is the Wind from TW Hydrae?}",
      journal = {\apj},
         year = 2007,
        month = jan,
       volume = {655},
       number = {1},
        pages = {345-350},
          doi = {10.1086/508770},
archivePrefix = {arXiv},
       eprint = {astro-ph/0609239},
 primaryClass = {astro-ph},
       adsurl = {https://ui.adsabs.harvard.edu/abs/2007ApJ...655..345J}
}

@ARTICLE{suzan_edwards2003,
       author = {{Edwards}, Suzan and {Fischer}, William and {Kwan}, John and {Hillenbrand}, Lynne and {Dupree}, A.~K.},
        title = "{He I {\ensuremath{\lambda}}10830 as a Probe of Winds in Accreting Young Stars}",
      journal = {\apjl},
         year = 2003,
        month = dec,
       volume = {599},
       number = {1},
        pages = {L41-L44},
          doi = {10.1086/381077},
archivePrefix = {arXiv},
       eprint = {astro-ph/0311289},
 primaryClass = {astro-ph},
       adsurl = {https://ui.adsabs.harvard.edu/abs/2003ApJ...599L..41E}
}

@ARTICLE{justyn_campbellwhite2023b,
       author = {{Campbell-White}, Justyn and {Manara}, Carlo F. and {Benisty}, Myriam and {et al.}},
      journal = {\apj},
         year = 2023,
       volume = {956},
       number = {1},
        pages = {25},
          url = {https://iopscience.iop.org/article/10.3847/1538-4357/acf0c0}
}

@ARTICLE{bayo2012,
       author = {{Bayo}, A. and {Barrado}, D. and {Hu{\'e}lamo}, N. and {et al.}},
      journal = {\aap},
         year = 2012,
       volume = {547},
        pages = {A80},
       adsurl = {https://www.aanda.org/articles/aa/full_html/2012/11/aa19374-12/aa19374-12.html}
}

@ARTICLE{a_siciliaaguilar2012,
       author = {{Sicilia-Aguilar}, A. and {K{\'o}sp{\'a}l}, {\'A}. and {Setiawan}, J. and {et al.}},
      journal = {\aap},
         year = 2012,
       volume = {544},
        pages = {A93},
          url = {https://ui.adsabs.harvard.edu/abs/2012A&A...544A..93S}
}

@ARTICLE{stauffer2014,
       author = {{Stauffer}, John and {Cody}, Ann Marie and {Baglin}, Annie and {et al.}},
      journal = {\aj},
         year = 2014,
       volume = {147},
       number = {4},
        pages = {83},
          url = {https://iopscience.iop.org/article/10.1088/0004-6256/147/4/83}
}

@ARTICLE{barrado2002a,
       author = {{Barrado y Navascu{\'e}s}, D. and {Zapatero Osorio}, M.~R. and {Mart{\'\i}n}, E.~L. and {et al.}},
      journal = {\aap},
         year = 2002,
       volume = {393},
        pages = {L85-L88},
          url = {https://ui.adsabs.harvard.edu/abs/2002A&A...393L..85B},
}

@ARTICLE{a_siciliaaguilar2020b,
       author = {{Sicilia-Aguilar}, A. and {Bouvier}, J. and {Dougados}, C. and {et al.}},
      journal = {\aap},
         year = 2020,
       volume = {643},
        pages = {A29},
          url = {https://www.aanda.org/articles/aa/full_html/2020/11/aa38489-20/aa38489-20.html}
}

@ARTICLE{m_tafalla2015b,
       author = {{Tafalla}, M. and {Hacar}, A.},
      journal = {\aap},
         year = 2015,
       volume = {574},
        pages = {A104},
          url = {https://www.aanda.org/articles/aa/full_html/2015/02/aa24576-14/aa24576-14.html}
}

@INPROCEEDINGS{p_andre2014,
       author = {{Andr{\'e}}, P. and {Di Francesco}, J. and {Ward-Thompson}, D. and {et al.}},
    booktitle = {Protostars and Planets VI},
         year = 2014,
        pages = {27-51},
          url = {https://muse.jhu.edu/chapter/1386878}
}

@ARTICLE{a_hacar2017b,
       author = {{Hacar}, A. and {Tafalla}, M. and {Alves}, J.},
      journal = {\aap},
         year = 2017,
       volume = {606},
        pages = {A123},
          url = {https://www.aanda.org/articles/aa/full_html/2017/10/aa30348-16/aa30348-16.html}
}

\end{document}